\documentclass[aps,prc,twocolumn,superscriptaddress,preprintnumbers,amsmath,amssymb,floatfix,showkeys,nofootinbib,linenumbers]{revtex4}

\usepackage{graphicx, fink}
\usepackage{epsfig}
\usepackage{dcolumn}
\usepackage{bm}
\usepackage{longtable}
\usepackage{color}

\usepackage{float}
\usepackage{comment}
\usepackage{amsmath}
\usepackage{amsfonts}
\usepackage{xspace}
\usepackage{multirow}
\usepackage{nicefrac}
\usepackage[normalem]{ulem}
\usepackage{graphicx}

\usepackage{array}

\makeatletter
\g@addto@macro\bfseries{\boldmath}
\makeatother

\usepackage{tikz}

\begin{document}

\title{Physics Opportunities with Meson Beams for EIC}

\author{\mbox{William~J.~Briscoe}}
\affiliation{The George Washington University, Washington DC 20052, USA}
\author{\mbox{Michael~D\"oring}}
\affiliation{The George Washington University, Washington DC 20052, USA}
\author{\mbox{Helmut~Haberzettl}}
\affiliation{The George Washington University, Washington DC 20052, USA}
\author{\mbox{D.~Mark~Manley}}
\affiliation{Kent State University, Kent, OH 44242, USA}
\author{\mbox{Megumi Naruki}}
\affiliation{Kyoto University, Kyoto 606-8502, Japan}
\author{\mbox{Greg~Smith}}
\affiliation{Thomas Jefferson National Accelerator Facility, Newport 
	News, Virginia 23606, USA}
\author{\mbox{Igor~Strakovsky}}
\affiliation{The George Washington University, Washington DC 20052, USA}
\author{\mbox{Eric~S.~Swanson}}
\affiliation{University of Pittsburgh, Pittsburgh, PA 15260, USA}

\date{\today}

\begin{abstract}
\textbf{Abstract:} Over the past two decades, meson photo- and electroproduction data of unprecedented quality and quantity have been measured at electromagnetic facilities worldwide. By contrast, the meson-beam data for the same hadronic final states are mostly outdated and largely of poor quality, or even non-existent, and thus provide inadequate input to help interpret, analyze, and exploit the full potential of the new electromagnetic data. To reap the full benefit of the high-precision electromagnetic data, new high-statistics data from measurements with meson beams, with good angle and energy coverage for a wide range of reactions, are critically needed to advance our knowledge in baryon and meson spectroscopy and other related areas of hadron physics. To address this situation, a state-of-the-art meson-beam facility needs to be constructed. The present letter summarizes unresolved issues in hadron physics and outlines the vast 
opportunities and advances that only become possible with such a facility.
\end{abstract}

\maketitle

\textbf{Introduction}: The CM energy range up to 2.5~GeV is rich in opportunities for physics with pion and kaon beams to study baryon and meson spectroscopy questions complementary to the electromagnetic programs underway at electromagnetic facilities. The White Paper~\cite{Briscoe:2015qia} highlights some of these opportunities and describes how facilities with high-energy and high-intensity meson beams can contribute to a full understanding of the high-quality data now coming from electromagnetic facilities. We emphasize that what we advocate here is not a competing effort, but an experimental program that provides the hadronic complement of the ongoing electromagnetic program, to furnish the common ground for better and more reliable phenomenological and theoretical analyses based on high-quality data.  A number of the topics mentioned in the White Paper are addressed in the summary of the recent DNP Town Meeting on QCD and Hadron Physics~\cite{Brodsky:2015aia}, which notes (on page 28) that meson beams are being considered.

The physics case for our program is aligned with \textit{Reaching for the Horizon: Long Range Plan for Nuclear Science}~\cite{Geesaman:2015fha}: \textit{...a better understanding of the role of strange quarks became an important priority.} Knowledge of the hyperon spectrum is an important component of this proposal. Overall, our knowledge of the hyperon spectrum is very poor; \textit{e.g.}, the empirical knowledge of the low-lying spectra of $\Lambda$ and $\Sigma$ hyperons remains very poor in comparison with that of the nucleon and, in the case of the $\Xi$ hyperons, extremely poor. The structure of hyperon resonances cannot be understood without empirical determination of their pole positions and decays, which is the goal of the proposed experiments. The determination of the strange hyperon spectra in combination with the current measurements of the spectra of the charm and beauty hyperons at the LHCb experiment at CERN allows for a clearer 
understanding of soft QCD matter and the approach to heavy quark symmetry~\cite{Isgur:1991wq}.

\textbf{Opportunities with Meson Beams}: Looking further down the road, we plan to take part in the future electron-ion collider (EIC) following the recommendation of the new Long-Range Plan \textit{Reaching for the Horizon: Long Range Plan for Nuclear Science}~\cite{Geesaman:2015fha}: \textit{RECOMMENDATION~III: We recommend a high-energy high-luminosity polarized EIC as the highest priority for new facility construction following the completion of FRIB.}

An EIC will likely be one of the future large accelerator facilities for high-energy and nuclear physics~\cite{AbdulKhalek:2021gbh}. 
The creation of a state-of-the-art hadron physics complex to study QCD at the deepest level with an EIC would provide the unique infrastructure for a meson-beam facility to complete our picture of the hadron spectrum of QCD at the same time. In addition, the presence of an electron beam at an EIC facility would permit a unique opportunity to measure the pion EM form factor directly using an electron-pion collider, although such a facility would present tremendous technical challenges. This is useful because the pion form factor serves as a paradigm for nonperturbative hadronic structure and is associated with chiral dynamics, gauge invariance, and perturbative QCD in a nontrivial fashion. Current methods to extract the form factor rely on extrapolation to the pion $t$-channel pole. Unfortunately, this procedure is not without ambiguity and remains controversial to this day.

The form factor is especially relevant in light of a more recent controversy concerning its approach to the expected perturbative QCD behavior. This issue is of fundamental importance since it questions the existence of perturbative QCD for exclusive 
processes~\cite{Gorchtein:2011vf}. Unfortunately, the experimental situation is confused, with the BaBar and Belle Collaborations obtaining results for $Q^2 |F_\pi(Q^2)|$ that appear to be in conflict~\cite{Bevan:2014iga}.

\textbf{Spectroscopy of Hyperon Resonances}: Most of our current knowledge about the bound states of three light quarks has come from partial-wave analyses (PWAs) of $\pi N\to\pi N$ scattering. Measurements of $\pi N$ elastic scattering are mandatory for determining absolute $\pi N$ branching ratios. Without such information, it is likewise impossible to determine absolute branching ratios for other decay channels. A summary of resonance properties (pole positions (masses and widths), branching ratios, helicity couplings to $\gamma N$, \textit{etc.}) is provided in the \textit{Review of Particle Physics} (RPP)~\cite{Zyla:2020zbs}.

The information on resonance properties obtained from analyses of experimental data provides fundamental information about QCD in the nonperturbative region.  A variety of quark models have been used to interpret these results.  Dyson-Schwinger approaches provide a picture of baryons in terms of quarks and gluons by incorporating dynamical chiral symmetry.  Results from lattice gauge theory calculations are constantly improving and are therefore becoming more relevant to experiment.

A comparison of the experimental results and models led to the well-known conundrum known as the ``missing resonances'' problem~\cite{Koniuk:1979vw}. Put simply, the models (and lattice-gauge 
calculations) predict far more states than are observed experimentally.  These missing resonances, however, do not appear at all in the quark-diquark model. The reason for this, it is conjectured, is their weak coupling to the $\pi N$ channel, which supplies the bulk of our 
information about baryonic resonance states.  A desire to test this hypothesis by looking for resonances in reactions that do not involve $\pi N$ in either the initial or final state was a major reason behind the construction of Hall~B and the CLAS facility at JLab.

The world data on $\pi N\to \eta N,\,K\Lambda, \,K\Sigma$ were collected and date back to more than 20 or 30 years ago.  In many cases, systematic uncertainties were not reported (separately from statistical uncertainties), and in many cases it is known that systematic uncertainties were underestimated.  These problems of pion-induced reaction data have led to the emergence of many different analyses that claim a different resonance content. While many analyses agree on the 4-star resonances that are visible in elastic $\pi N$ scattering~\cite{Arndt:2006bf}, there is no conclusive agreement on resonances that couple only weakly to the $\pi N$ channel.

Unlike in the cases described above, kaon beams are crucial to provide the data needed to identify and characterize the properties of hyperon resonances. The masses and widths of the lowest $\Lambda$ and $\Sigma$ baryons were determined mainly with kaon-beam experiments in the 
1970s~\cite{Zyla:2020zbs}. First determinations of pole positions, for instance for $\Lambda(1520)$, were obtained only recently~\cite{Qiang:2010ve}. An intense kaon beam would open a window of opportunity not only to locate missing resonances but also to establish properties including decay channels systematically for higher excited states.

In summary, better data from hadron-induced reactions will significantly contribute to answer the same fundamental questions that originally motivated the photoproduction program: the missing resonance problem, amplitudes for comparison with \textit{ab initio} calculations, and 
low-energy precision physics. A program with hadron beams provides complementary information with large impact in the extraction of the amplitudes from observables. With a more precise knowledge of the amplitude it is expected that much-debated concepts, such as the 
aforementioned multiquark hypotheses, hadronic molecules, hybrid states, chiral symmetry restoration, chiral solitons, or string models, can be confirmed or ruled out.

\textbf{Meson Spectroscopy}: Although it was light hadron spectroscopy 
that led the way to the discovery of color degrees of freedom and Quantum 
Chromodynamics, much of the field remains poorly understood, both 
theoretically and experimentally~\cite{Pennington:2014dwa}.  The 
availability of pion and kaon beams provide an important opportunity to 
improve this situation. Experimentally, meson spectroscopy can be 
investigated by using PWAs to determine quantum numbers from the angular 
distributions of final-state particle distributions.  Such methods will 
be used to analyze data from future measurements at CLAS12.  Pion beams 
with CM energies up to 5~GeV should be adequate for a complementary 
program.  Such energies correspond to beam momenta of about 13~GeV/$c$. We 
note that meson beam experiments may not be ideal for the study of meson 
resonances; however, this approach has been taken at BNL and COMPASS and 
a closely related one is being pursued by the GlueX Collaboration. We 
therefore briefly review some of the open issues in light meson 
spectroscopy in this section.

\textbf{What is Needed for Hadron-Induced Reactions}: These measurements 
also have the potential to observe dozens of predicted (but heretofore 
unobserved) states and to establish the quantum numbers of already 
observed hyperon resonances listed in the RPP~\cite{Zyla:2020zbs}. 
Interesting puzzles exist for RPP-listed excited hyperons that do not 
fit into any of the low-lying excited multiplets, and these need to be 
further revisited and investigated. Excited $\Xi$s, for instance, are 
very poorly known. Establishing and discovering new states is important, 
in particular, for determination of the multiplet structure of excited 
baryons.

Our expertise says that is doable at JLab. The use 3-10~GeV protons 
from pre-booster or booster (which will be busy several minutes a day) 
to produce secondary charged meson beams will increase the efficiency 
of the EIC facility.  One can roughly estimate that it will cost 
about 10\% of the EIC cost. 

\textbf{How it May Work}: 
What would be required to expand the scope of the EIC to include 
the production of secondary beams of pions and kaons, and what secondary 
fluxes might we expect? It's useful to recall the projects that were 
proposed (both successfully and unsuccessfully) in the past to provide 
such beams. 

\textbf{History}: Back in the 1980s, there were two proposals for kaon 
factories that failed to secure funding. The TRIUMF KAON 
factory~\cite{Craddock:1986nw,TRIUMF1,TRIUMF2} proposed to add an alternating 
series of three storage rings and two fast-cycling synchrotrons to the 
existing 150~$\mu$A, 0.5~GeV CW TRIUMF cyclotron. It would have provided 
100~$\mu$A of 30~GeV protons (3~MW) to an experimental area with four 
production targets to provide a range of secondary meson beams, as well as 
a dedicated line for neutrino physics, for Cdn \$700M (1986 dollars), or 
about \$1.1 B in today's (US) dollars. At the same time, a competing design 
at LAMPF called the Advanced Hadron Facility proposed to add a 1.2~GeV 
superconducting booster linac, which would have accepted 100~$\mu$A from 
the existing 0.8~GeV linac. The booster linac would have fed a 2~GeV 
compressor ring, followed by a 15~GeV booster ring and finally a 60~GeV 
main ring. This configuration allowed extraction of a short duty-factor 
neutrino beam and a high duty-factor beam for production of secondary 
particles. Both of these were (planned to be) MW-scale Kaon factories. 
Additionally, there was the the LAMPF II (AHF) proposal which described 
the 45-GeV version~\cite{AHF}.

BNL successfully operated kaon beam lines at the 30~GeV 
AGS~\cite{Littenberg:1983vn} for many years. Typical proton intensities of 
$6\times 10^{12}$ protons per second were achieved with a 43\% duty factor 
(DF), corresponding to almost 100~kW of average beam power. 

Today the highly successful J-PARC facility~\cite{JPARC,Sato:2010zzb,
Takahashi:2011zzf} in Japan is the only Kaon factory in the world.  It is a 
several-hundred kW proton accelerator with an extensive array of secondary 
kaon channels.

\textbf{The EIC Booster}: Several designs are under consideration for the 
EIC booster. Here we consider just one of these 
designs~\cite{Nissen:2018gdz}. It envisions an 8~GeV (figure-8 shaped) booster 
of circumference 313.5~m producing a $\sim1/4$ circumference, 260~ns-long spill 
every 2~s  with a parabolic intensity distribution (the booster synchrotron 
magnets take 1~s to ramp up and 1~s to ramp down). The design intensity of 
$1 \times 10^{12}$ protons (160~nC) per spill corresponds to only about 0.6~kW, 
2 orders of magnitude below the BNL AGS when it was serving kaon beamlines, and 
3 orders of magnitude less than J-PARC~\cite{JPARC}. The time-averaged current 
is only 80~nA.

It seems possible to consider flattening the distribution to 840~ns, or 
increasing the intensity a factor of 2-3 for modest additional investment. Much 
more expensive would be to increase the intensity significantly, or to 
slow-extract over (for example) 2~s every 3~s. While these changes would offer 
significant advantages, we take the position for the remainder of this manuscript 
that the booster parameters are frozen as described in Ref.~\cite{Nissen:2018gdz}, 
and then look to see what would need to be done on top of the existing design to 
satisfy the requirements for $\pi$ and K beams outlined above. Since the EIC 
booster would only inject to the main ring a small fraction of the time each day, 
we assume it could be made available for $\pi$/K physics the rest of the time. 

\textbf{Stretcher}: The nominal booster design has a duty factor of only 0.00005\%, 
which makes it completely unsuitable for the physics experiments described earlier in 
this manuscript. Most experiments rely on detectors that are limited to 
instantaneous rates below a few MHz, so it's better to have the same number of 
beam particles delivered with a lower instantaneous rate over a longer period of 
time than all at once in a short period of time.  As noted, for example, 
in Ref.~\cite{Arvieux}, ``\textit{In coincidence measurements the number of true events 
is proportional to the intensity but the number of random coincidence events 
(background) goes like the square of the peak current}". Without improving the 
duty factor of the nominal EIC booster there is no reason to proceed any further. 
Of course if the booster itself could be modified to provide slow extraction, a 
possibility mentioned above, that would be the most cost-effective solution to the 
duty-factor problem. 

There is a solution to this seemingly fatal problem that does not involve 
making changes to the nominal booster design: adding a stretcher ring (see, for 
example, Ref.~\cite{Tomizawa:2018koh}). The nominal 8~GeV/$c$ booster corresponds 
to a magnetic rigidity of 26.67~T-m. There are many stretcher configurations that 
could be used to provide nearly CW 100\% duty factor beam in this situation. The 
simplest and most basic would be to extract the booster beam in a direction 
different than used to fill the EIC main ring, and fill exactly one turn of a 
stretcher ring over the 260~ns booster spill. That implies a booster ring with a 
radius of 12.3~m, and 39 one-meter-long dipole magnets of 4.3~T (fixed field) if 
half the circumference of the ring consisted of dipoles. While this is feasible, 
it would be more sensible to (for example) double the stretcher diameter, fill 
only half its circumference every spill, and halve the field to 2.16~T so 
resistive magnets could be used. In this mode, and with some flexibility in the 
bunch length, the planned booster tunnel would be about the right circumference 
to house the stretcher magnets. The stretcher magnet design could be the the same 
or similar to that of the booster magnets. 

Either way the beam could be extracted essentially CW with a good duty factor using 
the common technique of resonant slow extraction (see, for 
example~\cite{resonantslowextraction}) while the booster magnets are ramping up and 
down. In this way, a time-average of $5 \times 10^{11}$ protons/s could be 
slow-extracted from the stretcher ring into a transfer line to a production target 
for secondary particles ($\pi$/K) in a dedicated hadron hall. Now we look to see 
what $\pi$/K fluxes might be achieved with a CW proton beam of 80~nA.

\textbf{Pion and Kaon Intensities}: Given the conditions just established (80~nA CW, 
8~GeV/$c$ protons) we can estimate the fluxes of pions and kaons as a function of 
secondary particle momentum and secondary channel production angle. We first use 
the parametrizations of Sanford and Wang (SW) to estimate the number of 
pions~\cite{SW1} and kaons~\cite{SW2} generated at the production target per GeV/$c$ 
per steradian per interacting proton. Then we use those results as input to the 
procedure described by Yamamoto~\cite{Yamamoto:1981zr}, which includes the effects 
of secondary particle production efficiency and nuclear absorption cross sections 
of the protons on different production targets, as well as solid angle, momentum 
acceptance, and decay factors for typical secondary-particle channels. This tells 
us what to expect for $\pi^\pm$ and K$^\pm$ fluxes at the end of the secondary 
particle beamline, again as a function of production angle, for a given production 
target and secondary-particle beamline. Yamamoto described the design of the KEK K2 
channel for $1 \times 10^{12}$ protons at 12~GeV, which are similar conditions to the ones
we're considering here.

It must be noted that the SW parametrizations \cite{SW1,SW2} used proton data 
between 10 and 30~GeV. We are outside this range for the 8~GeV case we're 
considering here, which decreases  the reliability of our predictions somewhat. For 
the production target and secondary beam channel  we consider Yamamoto's 
design~\cite{Yamamoto:1981zr} as the default: a 6~cm Pt production target, 30~m 
channel with 3.125~msr and $\Delta p/p$=2\%. This is a realistic design for our situation, 
and also makes it easier to benchmark our calculations against  his.

\begin{figure*}[htb!]
\centering
\begin{minipage}{1\textwidth}
\includegraphics[width=1.0\textwidth]{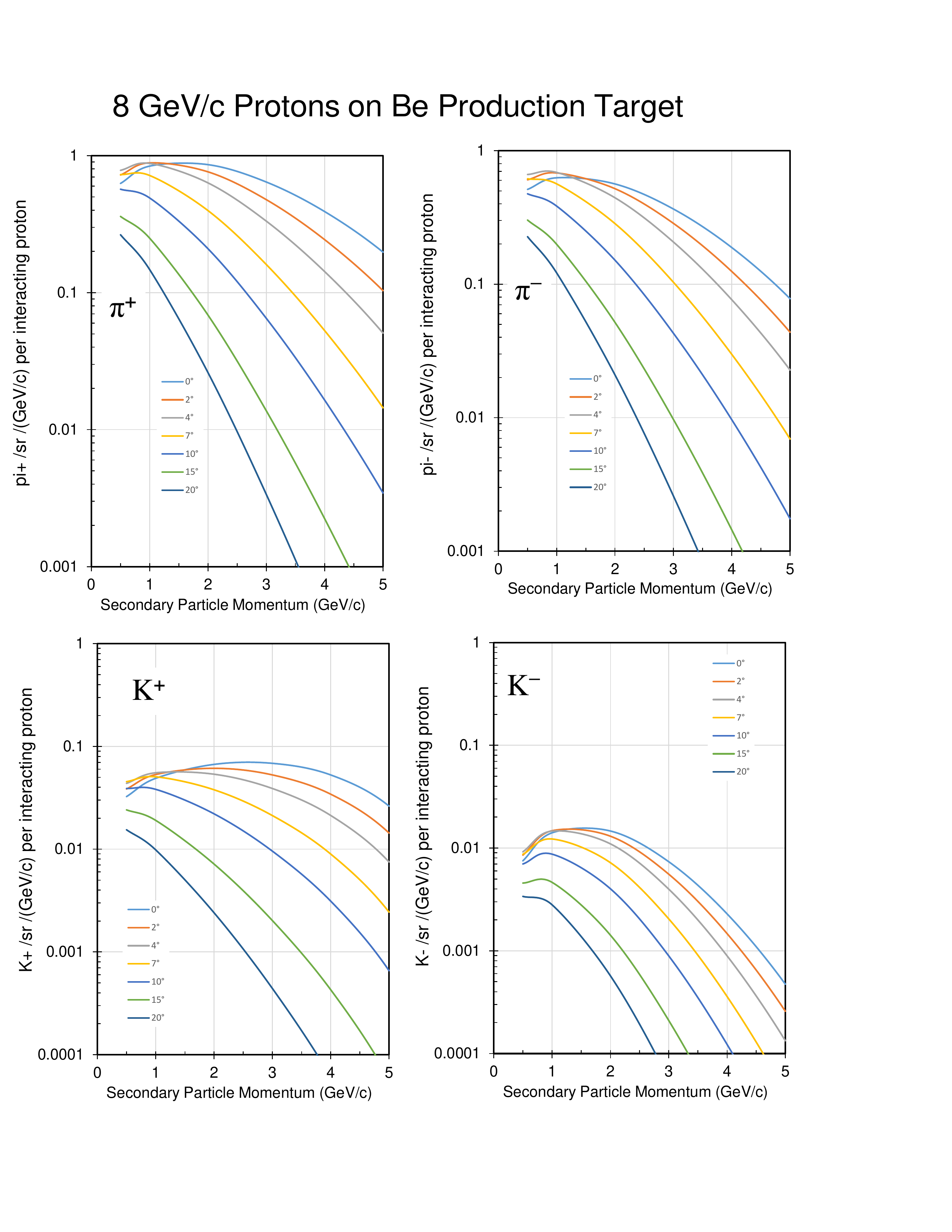}
\vspace*{-2.5cm}

\caption{Pion and Kaon production rates on beryllium per steradian per GeV/$c$ per 
	interacting proton for $5 \times 10^{11}$ protons/s CW at 8~GeV. The 
	legend denotes different assumptions for the production angle.}
\label{fig:SW1}
\end{minipage}
\end{figure*}

Figure~\ref{fig:SW1} shows the predictions of $\pi^\pm$ and K$^\pm$ production on 
beryllium per steradian per GeV/$c$ per interacting proton for $5 \times 10^{11}$ 
protons/s CW at 8~GeV from the EIC booster and stretcher, based on the 
phenomonological fits of SW~\cite{SW1,SW2}. Near production angles of $0^\circ$, 
and for $\pi$/K momenta near 3~GeV/$c$,  $\pi^\pm$ production rates are just under 
1/sr/(GeV/$c$) per interacting proton, with K$^+$ rates about 30~times less than 
that, and K$^-$ rates another factor of 5 below that.

Much more interesting is to use this information to predict what fluxes of  
secondary particles could be expected for a specific choice of production 
target and realistic secondary beamline. Following the work of 
Yamamoto~\cite{Yamamoto:1981zr}, the $\pi$K yields $Y$  can be determined from 
the SW rates $d^2 N /(d \Omega dp) = \left( 1/\sigma_a \right) d^2 \sigma /(d 
\Omega dp)$, the proton flux $F_i$ on the production target, the production 
efficiency $\eta$ for secondary particles per incident proton, the nuclear 
absorption cross section $\sigma_a$ associated with the protons on the production 
target, as well as the secondary channel characteristics: solid angle $d \Omega$, 
momentum bite $\Delta p /p$, momentum $p$, and decay factor $d$ according to
\begin{equation}
	Y=F_i \,  \frac{\eta}{\sigma_a} \frac{d^2\sigma}{d\Omega dp} \Delta 
	\Omega \, (\Delta p/p)\, p \, d.
    \label{eq:yields}
\end{equation}
We take an incident proton flux $F_i = 5 \times 10^{11}$ protons/s CW at 8~GeV, 
a 6~cm Pt production target for which $\sigma_a=1798$~mb, $\eta=0.365$. We scale 
$\sigma_a$ and $\eta$ from the SW rates that were for a Be production target 
($\sigma_a = 227$~mb and $\eta=0.14$). The secondary channel is 30~m long with 
3.125~msr and $\Delta p/p$=2\%. The decay factor $d$ is the probability the secondary 
particle of mass $M$, mean life $\tau$, and momentum $p$  survives a distance 
$x = 30$~m:
\begin{equation}
    d(x)=\exp{\left(-Mx/(p \;c\tau)\right)}.
\end{equation} 
Following Yamamoto~\cite{Yamamoto:1981zr}, we determine the production efficiency 
from the absorption lengths in Ref.~\cite{Denisov:1973zv} for protons, pions, and 
kaons. Although the absorption cross sections tabulated in Ref.~\cite{Denisov:1973zv} 
are at higher momenta than in our application, they are reasonably flat. This is 
also apparent from the corresponding plots of total cross sections (Figs.\ 
51.6-51.9) in the RPP~\cite{Zyla:2020zbs}.  We determine the nuclear 
absorption cross section for the protons in the production target from 
Ref.~\cite{Denisov:1973zv} and note that Yamamoto~\cite{Yamamoto:1981zr} points 
out that it scales like $A^{2/3}$.  For forward-angle production, 3~GeV/$c$ pion 
fluxes of about 10~MHz are possible. Similarly, K$^+$ fluxes are an order of 
magnitude less, and K$^-$ fluxes are another order of magnitude below that. 

These modest kaon fluxes would not inspire rare-decay experiments; however, they 
do seem well-suited to the physics program described earlier in this manuscript. 
At these $\pi$/K rates, the beam can be directly counted, which offers important 
benefits to experimenters working with these beams. Although the low power 
associated with the proton beam (0.6~kW) is the root cause of the 
lower-than-ideal $\pi$/K rates, it has a silver lining: it makes it much easier 
(and less expensive) to deal with the shielding, rad-hardening of magnets, 
activation issues, and cooling of the production target and surrounding region 
than has been the case with higher power beams~\cite{Densham:2010fva,Agari:2012gj,
Agari:2005zz}.

Secondary particle channels have been designed (see, for example~\cite{Yamamoto:1981zr,
Pile:1992vk,Doornbos:2000hb}) at a number of different facilities that would serve 
the application described here as well. They employ one, usually two Wien 
filters~\cite{Ieiri:2013cpa} (crossed \textbf{E} and \textbf{B} field regions) to 
select either $\pi$ or K beams. 

\begin{figure*}[htb!]
\centering
\begin{minipage}{1.0\textwidth}
\includegraphics[width=1.0\textwidth]{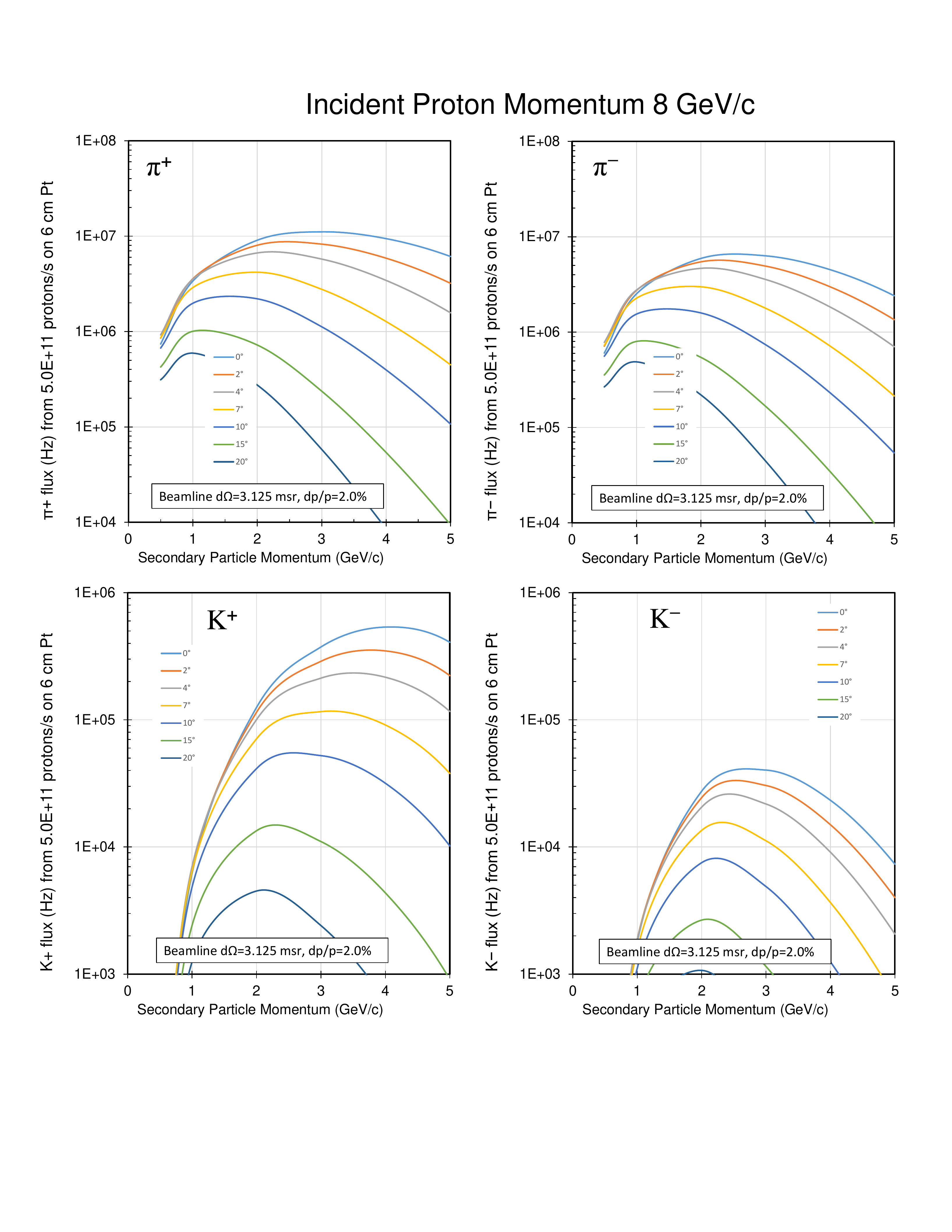}

\vspace*{-4cm}
\caption{Pion and Kaon  rates that could be expected at the end of a 
	30~m-long channel with a solid angle of 3.125~msr and a momentum 
	bite of 2\% $\Delta p/p$. The production target is assumed to be 6~cm Pt. 
	Decay along the length of the channel, as well as secondary 
	particle production efficiency and nuclear absorption cross
	sections of the protons on the Pt target are also factored in 
	to this prediction.  The legend denotes different assumptions 
	for the production angle.}
\label{fig:SW2}
\end{minipage}
\end{figure*}

\textbf{Summary}: The goals of current EM facilities would benefit 
greatly from having hadron-beam data of a quality similar to that of 
EM data. To this end, it is commonly recognized that a vigorous U.S.\ 
program in hadronic physics requires a modern facility with pion and 
kaon beams. A pion beam and a facility in which $\pi N$ elastic 
scattering and the reactions $\pi^-p\to K^0\Lambda$, $\pi^-p\to 
K^0\Sigma^0$, $\pi^-p\to K^+\Sigma^-$, and $\pi^+p\to K^+\Sigma^+$ 
can be measured in a complete experiment with high precision would 
be very useful.  Full solid angle coverage is required to study 
inelastic reactions such as $\pi^-p\to\eta n$, $\pi^+p\to\pi^0\pi^+p$, 
or strangeness production (among many other reactions). Such a 
facility ideally should be able to allow baryon spectroscopy 
measurements up to center-of-mass energies $W$ of about 2.5~GeV, 
which would require pion beams with momenta up to about 2.85~GeV/$c$. 
The 2~GeV/$c$ pion beam at J-PARC will allow baryon spectroscopy 
measurements up to $W \approx 2150$~MeV.

The White Paper~\cite{Briscoe:2015qia} outlined some of the physics programs that could be advanced with a hadron-beam facility. These include studies of baryon spectroscopy, particularly the search for ``missing resonances'' with hadronic beam data that would be analyzed together with photo- and electroproduction data using modern coupled-channel analysis methods. A hadron beam facility would also advance hyperon spectroscopy and the study of strangeness in nuclear 
and hadronic physics.

At the end of the White Paper~\cite{Briscoe:2015qia}, there is a list of endorsers who have expressed support for the initiative described herein: 135 researchers from 77 institutes representing 20 countries around the world.

\textbf{Acknowledgment:}
We thank the numerous colleagues who have contributed either directly or indirectly to this work. This material is based upon work supported by the U.S. Department of Energy, Office of Science, Office of Nuclear Physics under Contracts No. DE--SC0016583, DE--SC0016582, and DE--SC0014323.


\end{document}